\documentstyle[12pt]{article}

\def\b{\beta}

\def\d{\delta}

\def\m{\mu}

\def\q{\theta}

\def\s{\sigma}

\def\cu{{\cal U}}

\def\pro{\propto}
\def\la{\left}
\def\ra{\right}
\def\pa{\partial}

\def\abs#1{\left| #1\right|}

\def\Hat#1{\rlap{\kern.10em$\widehat{\phantom G}$}#1}
\def\HAt#1{\rlap{\kern.05em$\widehat{\phantom G}$}#1}

\def\cap#1{\rlap{\kern.1em$\widehat{\phantom{G\vrule height.8em}}$}#1{}}
\def\Cap#1{\rlap{\kern.05em$\widehat{\phantom{G\vrule height.8em}}$}#1{}}

\def\ev#1{\left\langle #1\right\rangle}

\catcode`@=11
\def\underline#1{\relax\ifmmode\@@underline#1\else
        $\@@underline{\hbox{#1}}$\relax\fi}
\catcode`@=12

\def\PRL{Phys. Rev. Lett.\ }

\def\PRD{Phys. Rev. D\ }

\def\NPB{Nucl. Phys. B\ }

\def\CMP{Comm. Math. Phys.\ }

\def\ba{\begin{array}}
\def\ea{\end{array}}
\def\be{\begin{equation}}
\def\ee{\end{equation}}
\def\bdm{\begin{displaymath}}
\def\edm{\end{displaymath}}
\def\bea{\begin{eqnarray}}
\def\eea{\end{eqnarray}}

\def\by{\over}
\def\lb{\label}
\def\bl#1{(\ref{#1})}

\def\vsp{\\ \vglue 0.2in}
\def\au{{\it Department of Physics\\
          University of Arizona, Tucson, AZ 85721}}

\def\cu{{\it Department of Physics\\
          University of Colorado, Boulder, CO 80309}}

\def\bt#1#2#3#4#5
        {\begin{titlepage}
         \centerline{\hfill {#1}}
         \centerline{\hfill {#2}}
         \centerline{\hfill {#3}}
         \begin{center}\vskip .2in
         {\large\bf {#4}}\\[.3in]
         {\bf {#5}}\\[.3in]
         {\bf ABSTRACT}\\
         \end{center}
         \begin{quotation}}
\def\et
         {\end{quotation}
          \end{titlepage}
          \setcounter{page}{2}}

\newcounter{sxn}

\newcounter{axn}

\def\br{\begin{thebibliography}{abc}}
\def\er{\end{thebibliography}}
\def\rf{\bibitem}
\def\fr#1{\cite{#1}}
\begin{document}

\bt{COLO-HEP/313, AZPH-TH/93-10}{hep-th/9304079}{April 1993}{Anomaly
matching for the QCD string}{B. S. Balakrishna \vsp
\footnote{Address for correspondence. Email:
bala@haggis.colorado.edu}\cu \vsp and \vsp \au}

A criterion to be satisfied by a string theory of QCD is formulated in
the ultraviolet regime. It arises from the trace anomaly of the QCD
stress tensor computed using instantons. It is sensitive to asymptotic
freedom. It appears to be related to the trace anomaly of the QCD
string. Our current understanding of noncritical strings in physical
dimensions is limited, but remarkably, a formal treatment of the
bosonic string yields numerical agreement both in magnitude and sign
for the gauge group SU(2).

\et

It is a long held belief that QCD has a formulation based on string
theory. There are both experimental and theoretical reasons to expect
this. On the experimental side, the success of dual models in
describing some of the low energy phenomenon is quite
compelling\fr{rev}. Support from the theoretical side comes from 't
Hooft's $1/N$ analysis of SU($N$) QCD that has a topological expansion
like in string theory\fr{tht}. Wilson's lattice gauge theory in the
strong coupling expansion suggests area confinement and leads to a sum
of surfaces reminding us of strings\fr{wls}. Other reasons include a
picture of quark confinement via flux tubes that are expected to
behave like strings at least qualitatively.

It is not clear whether QCD has an exact formulation based on strings.
No satisfactory string model currently exists that could describe all
the phenomenon of strong interactions. The absence of massless
particles in the low energy spectrum of QCD, if not tachyons, makes it
difficult to relate it to the known string theories. Further, it is
not clear how asymptotic freedom that is an integral part of QCD could
be incorporated into the string picture. It could be that the simplest
possibility, the Nambu-Goto string, is not capable of describing QCD.
An analytical continuation of the confining phase to high temperature
suggests that this might be the case\fr{plc}. A string theory of QCD,
if one exists, is likely to be unconventional. It has escaped us
perhaps due to our limited understanding of noncritical strings in
physical dimensions.

In this letter, we do not address these problems. Rather, we look for
any common threads within the two pictures proceeding under the
assumption that some kind of a QCD string does exist at all distance
scales. Thanks to asymptotic freedom, this is expected to be easier in
the ultraviolet regime. Surprisingly, it is found possible to
formulate a short distance criterion to be satisfied by a string
theory of QCD. This arises from the trace anomaly of the QCD stress
tensor computed using instantons. Being proportional to the beta
function, it carries one important information we have on QCD, its
asymptotic freedom. A string candidate for QCD is expected to exhibit
this anomaly. It appears to be calculable in terms of the trace
anomaly of the string. Due to our insufficient understanding of
noncritical strings in physical dimensions, this is a difficult test
to be verified. Remarkably though, we find numerical agreement both in
magnitude and sign for pure SU(2) QCD in a string model treated
formally. Because this test appears to be sensitive to the trace
anomaly of the string, it is expected to severly limit the effective
world sheet degrees freedom in the ultraviolet regime.

Let us first concentrate on the trace anomaly of the QCD stress
tensor. We will be working in the Euclidean region and mostly
concerned with pure QCD. QCD Lagrangian in the absence of quark masses
is scale invariant and the only contribution to the trace $\q$ of the
stress tensor is from the anomaly. This is given in terms of the beta
function as\fr{qcd}
\be
\q ~=~ \b(g){\pa\by\pa g}L,
\ee
where $L$ is the QCD Lagrangian and $g$ is the strong coupling
constant. This follows from the following observation. A scaling of
the coordinates in a Green's function can be implemented by a scaling
of the cutoff parameter in a theory like QCD that has no dimensionful
coupling constants. But, as is well known for renormalizable theories,
the latter scaling can be compensated by a change in the coupling
constant and this introduces the beta function. The trace of the
stress tensor, being the divergence of the generator of scale
transformations, thus picks up an anomaly proportional to the beta
function. Our interest is in the vacuum expectation value of $\q$ in
the asymptotic regime, that is as $g\to 0$. More specifically, we are
interested in taking the $g\to 0$ limit of $\ev{\q}/F$ where $F$ is
the QCD free energy. This limit would have vanished if not for
instantons. It is well known that in the dilute gas approximation the
instanton contribution in SU($N$) QCD is\fr{clm}
\be
F ~\pro~ g^{-4N}{\rm exp}(-8\pi^2/g^2),
\ee
The desired limit is now easily taken,
\be
{\ev{\q}\by F} ~=~ \b(g){\pa\by\pa g}{\rm ln}F ~\to~ -{11\by 3}N,
\ee
making use of the one loop beta function of pure QCD,
\be
\b(g) ~=~ -{11N\by 48\pi^2}g^3.
\ee
Thus $\ev{\q}/F$ has a finite limit as $g\to 0$. It is enough to know
the one loop beta function to reach this limit. Dilute gas
approximation for instantons is likely to be sufficient to compute
this limit exactly.

Instanton contributions are negligibly small in the limit $g\to 0$
compared to any perturbative ones. But we retained only those in the
process of taking the limit. This is because our interest is in the
part of QCD that is of relevance to the string picture. Note that the
QCD string is a nonperturbative phenomenon expected to describe the
confining phase of QCD. Stringy features are thus going to be singular
at $g=0$. The leading contribution to this singular part is likely to
be from instantons. In the rest of this letter, this singular part is
referred to simply as QCD free energy.

We looked at $\ev{\q}$ divided by the free energy because that is more
interesting than $\ev{\q}$ itself. It has a natural meaning in the
string picture. Note the partition function of the string is expected
to give us not the QCD partition function but its free energy. Hence,
it is $\ev{\q}/F$ that has an interpretation as a world sheet
expectation value. It should agree with the expectation value of an
analogous object evaluated on the world sheet. Because this quantity
has a finite ultraviolet limit in QCD and is sensitive to asymptotic
freedom, it should be regarded as a strong test for the QCD string
candidates.

It is thus important to check whether there exist any string theories
exhibiting the same trace anomaly. Unfortunately, there are no known
consistent string candidates for QCD. For the purpose of illustrating
how this anomaly could arise in the string picture, we here address a
bosonic string governed by a two dimensional field theory of the space
time coordinates $X$ and the ghosts $b$ and $c$ defined on the world
sheet, namely a two dimensional sphere\fr{gsw}. For the usual action,
this is not a well defined theory since its BRST operator is not
nilpotent outside 26 dimensions\fr{brs}. However, we leave open the
possibility that there could be a suitable theory that cures this
problem. This is more so in the regime we are looking at since one
expects important corrections to the usual action that become relevant
at short distances\fr{plk}. A metric $h$ in the background governs the
geometry of the world sheet with an overall scale factor determining
the size. There is another metric on the world sheet induced by an
embedding into space time. Metric $h$ is expected to be correlated to
an effective value of this induced metric. It is hence a convenient
measure of the distance scale being probed. Note also that the
integration measure for the $X$ fields, a measure of the quantum
fluctuations of the world sheet, depends on $h$ through the norm
\be
\abs{\abs{\d X}}^2 ~=~ \int \sqrt{{\rm det}~h}\abs{\d X}^2,
\ee
where the integral is over the sphere. Thus, a scaling of the
coordinates could be implemented by a scaling of the metric $h\to
e^{\s}h$.

Important contributions to this scaling arise from the integration
measures $D_h(X)$, $D_h(b)$ and $D_h(c)$ of the $X$, $b$ and $c$
fields respectively. As is well known, these measures scale as
\be
D_{h}(X)D_{h}(b)D_{h}(c) ~\to~ D_{h}(X)D_{h}(b)D_{h}(c)~{\rm
exp}(-S_\s),
\ee
where
\be
S_\s ~=~ {26-D\by 48\pi}\int\la({1\by 2}\pa\s\pa\s + R\s +
m^2e^{\s}\ra).
\ee
Dependence on the metric $h$ is implicit in this expression. $D=4$ is
the space time dimension. $R$ is the scalar curvature of the sphere.
The number 26 comes from the ghosts $b$ and $c$, and $D$ is from the
coordinates $X$. In our case $\s$ is a constant. The $m^2$ term could
be dropped because it is proportional to the world sheet area expected
to be negligible in the ultraviolet limit. Now, using the following
well known theorem on the sphere,
\be
\int R ~=~ 8\pi,
\ee
we find that the measure contributes a factor
\be
{\rm exp}\la[-(26-D)\s/6\ra].
\ee
The exponent gives the trace anomaly of the world sheet stress tensor.
Because we have treated the background metric as a measure of the
distance scale being probed, it gets tied to the trace anomaly of the
space time stress tensor. It is to be compared with our result for
QCD. First, note that the scaling $h\to e^{\s}h$ corresponds to a
scaling of the coordinates by $e^{\s/2}$. Hence the exponent of
interest is $(26-D)/3$. Remarkably, in four dimensions, this exponent
coincides with our result $11N/3$ for pure SU(2) QCD.

It turns out that the above agreement for SU(2) is both in magnitude
and sign. To see this, let us relate the two anomalies given $f$ (the
world sheet integral of) the free energy of the string and $F$ that of
QCD. Recall that the string partition function ${\rm exp}(-f)$ is
expected to agree with the QCD free energy. In the presence of some
kind of a source, the relation is upto a constant
\be
f[x,h] ~=~ -{\rm ln}F[x,\m,g(\m)].
\ee
We have specified the dependence on the source locations collectively
as $x$. Also, we have included in $f$ a dependence on the background
metric $h$, and in $F$ a dependence on the renormalization point $\m$
and the running coupling constant $g(\m)$. Now, a scaling of $x$ can
be passed on to that of $h$ and $\m$ by dimensional analysis, and this
leads to
\be
f\la[x,e^{\s}h\ra] ~=~ -{\rm ln}F\la[x,e^{-\s/2}\m,g(\m)\ra]. \lb{zz}
\ee
Renormalizability of QCD is a statement that the QCD free energy is
$\m$ independent. Hence, a scaling of $\m$ can be compensated by a
change in the coupling constant $g$, that is
\be
F\la[x,e^{-\s/2}\m,g(\m)\ra] ~=~ F\la[x,\m,g\la(e^{\s/2}\m\ra)\ra].
\ee
Using this in \bl{zz} and going to infinitesimal $\s$, we get the
result
\be
h{\pa\by\pa h}f ~=~ -{1\by 2}\b(g){\pa\by\pa g}{\rm ln}F,
\ee
where the beta function is introduced through $\b(g)=\m dg(\m)/d\m$.
In deriving this, we have ignored any other dimensionful parameters
besides $h$ that could be present in $f$. But it is reasonable to
assume that those parameters become insignificant at short distances.
Dimensionless parameters if present do not interfere in our
discussion. Note that we do not have to change them as we did for
$g(\m)$.

As noted earlier, the right side of the above equation comes from the
trace anomaly of the QCD stress tensor. On the left side, we have (an
integral of) the trace of the world sheet stress tensor giving its
trace anomaly. Hence, this is an anomaly matching consistency check
for a string picture of QCD. It should be valid in the absence of
sources. We have already computed the two sides of this equation in
the ultraviolet limit. The right side is $11N/6$ in pure QCD. The left
side is $(26-D)/6$. In four dimensions, the two agree both in
magnitude and sign for the gauge group SU(2). This is remarkable given
that the two anomalies were computed from completely different
approaches. The agreement in sign is a pleasant surprise since the
sign is crucial for asymptotic freedom in QCD. The restriction to
SU(2) is understandable given that introducing an $N$ dependence on
the world sheet must involve some yet unknown physics. Fortunately for
SU(2) all the numerical factors combine to give agreement.

There are reasons to suspect that the numerical agreement could be
just a coincidence. We have ignored the contributions coming from the
action since we do not know what that constitutes at short distances.
Independence over the background metric has not been ensured. If it
were to become a dynamical variable, with its conformal factor
becoming a field variable on the world sheet, we would get a Liouville
field theory. The highly nontrival integration measure of this field
could modify our conclusions. The numerical agreement is only for
SU(2) and not for SU($N$) of higher $N$. But, whatever the reasons,
one can not escape the fact that an ultraviolet test of this kind is a
useful tool to uncover string physics from QCD.

Some of the problems may have natural solutions at short distances.
The action could be scale invariant or its variation negligible. The
background metric could be simply a classical solution for the induced
metric. Note that a rigid string\fr{plk} with a scale invariant action
involving the extrinsic curvature yields as a classical solution a
round metric with an overall scale undetermined. The overall scale may
play the role of a renormalization point so that independence over it
could be ensured in the presence of a running coupling constant. If we
were to have a Liouville field theory, it is possible that the nonzero
modes are suppressed at short distances. Extension to $N>2$ may
require an $N$ dependent theory, or a theory in a suitable phase may
introduce a factor $N$ into the scaling arguments.

There are also reasons to suspect that the numerical agreement could
be more than a coincidence. The existence of a finite ultraviolet
limit for the trace anomaly in the presence of instantons requires a
matching one in the string picture. The number 11 that appears in the
one loop beta function for QCD and also as a factor in the trace
anomaly of the string in four dimensions does not arise often in
computations. It is nontrivial that the rest of the factors combine
neatly to give agreement for SU(2). Topological results are needed in
both the cases, instantons in QCD and Gauss-Bonnet theorem on the
sphere. There are ghosts contributing to the anomaly in both the
pictures. It is thus reasonable to expect a relation of some kind for
a QCD string, if one exists.

Note that the trace anomaly $\pro 26-D$ strongly influences our
conclusions. Because this anomaly depends on the number of effective
degrees of freedom living on the world sheet, it is perhaps unlikely
that more of them are excited at short distances as argued in ref.
\fr{plc}. Asymptotic freedom is crucial as it is responsible for the
agreement in the signs. Our analysis is for pure QCD and having quark
degrees of freedom in the theory makes asymptotic freedom less severe.
It is interesting to note that the same behavior is observed in string
theory where the number $26-D$ decreases for more degrees of freedom.
For every quark, we expect a decrease by two units for the anomalies
to match.

The number $11N/3$ we have tried to match coming from the trace
anomaly is proportional to $N$. It is not clear how such an $N$
dependence could arise in the string picture based on 't Hooft's $1/N$
analysis. Perhaps this is because the latter is implicitly based on a
perturbative argument in the QCD coupling constant and one needs to
incorporate instanton effects into it before addressing this issue.
It is not surprising that instantons play a major role in our
discussion. The string picture is expected to hold in the confining
phase of QCD with the instantons possibly contributing some of the
nonperturbative effects. That instantons could have implications for
the QCD string is also apparant from the results of ref. \fr{aty}
which identifies the Yang-Mills instantons with the instantons of a
two dimensional theory taking values in the loop group. Clearly, there
is a lot to be learnt.

To summarize, it is shown in this letter that there exists a short
distance criterion to be satisfied by a string theory of QCD. This
arises from the trace anomaly of the QCD stress tensor in the presence
of instantons. Being proportional to the beta function, it is
sensitive to asymptotic freedom. It appears to be related to the well
known trace anomaly of string theory. Because of our limited
understanding of noncritical strings in physical dimensions, this is a
difficult test to be verified. However, a formal treatment of a string
model leads to remarkable numerical agreement between the two pictures
for the gauge group SU(2) both in magnitude and sign.

This work is supported by NSF grant PHY-9023257.

\br
\rf{rev}
See for instance P. D. Collins, ``Regge Theory and High Energy
Physics'', Cambridge University Press, Cambridge 1977.
\rf{tht}
G. 't Hooft, \NPB 72, 461 (1974).
\rf{wls}
K. G. Wilson, \PRD 10, 2445 (1974).
\rf{plc}
J. Polchinski, \PRL 68, 1267 (1992).
\rf{qcd}
J. C. Collins, A. Duncan and S. D. Joglekar, \PRD 16, 438 (1977) and
references therein.
\rf{clm}
See for instance S. Coleman, ``Aspects of Symmetry'', Cambridge
University Press, Cambridge 1985.
\rf{gsw}
See for instance M. B. Green, J. H. Schwartz and E. Witten,
``Superstring Theory'' Vol. I, Cambridge University Press, Cambridge
1987.
\rf{brs}
M. Kato and K. Ogawa, \NPB 212, 443 (1983).
\rf{plk}
A. M. Polyakov, \NPB 268, 406 (1986), and ``Gauge Fields and
Strings'', Harwood Academic Publishers, 1987.
\rf{aty}
M. F. Atiyah, \CMP 93, 437 (1984).
\er

\end{document}